\documentclass{ws-rv975x65}

\usepackage{newcent} 

\begin{document}
\setcounter{chapter}{0}

\long \def \blockcomment #1\endcomment{}
 
\chapter{YANG-MILLS FIELDS AND THE LATTICE}

\markboth{M. Creutz}{Yang-Mills fields and the lattice}

\author{Michael Creutz
}

\address{Physics Department, Brookhaven National Laboratory\\
Upton, NY 11973, USA\\
E-mail: creutz@bnl.gov}


\begin{abstract}
The Yang-Mills theory lies at the heart of our understanding of
elementary particle interactions.  For the strong nuclear forces, we
must understand this theory in the strong coupling regime.  The
primary technique for this is the lattice.  While basically an
ultraviolet regulator, the lattice avoids the use of a perturbative
expansion.  I discuss some of the historical circumstances that drove
us to this approach, which has had immense success, convincingly
demonstrating quark confinement and obtaining crucial properties of
the strong interactions from first principles.
\end{abstract}

\section{Introduction}

Originally motivated to extend the gauge theory of quantum
electrodynamics to include isospin, the Yang-Mills theory has become a
core ingredient of all modern theories of elementary particles.  With
the particular application to the strong interactions of quarks
interacting by exchanging non-Abelian gauge gluons, some rather unique
issues arise.  In particular, asymptotic freedom and dimensional
transmutation imply that low energy physics is controlled by large
effective coupling constants.  Long distance phenomena, such as chiral
symmetry breaking and quark confinement, lie outside the realm of
accessibility to the traditional Feynman diagram approach.  This drove
us to new approaches, amongst which the lattice has proven the most
successful.

This chapter is a personal reminiscence of how the lattice approach
was developed and grew to become the dominant approach to study
non-perturbative effects in quantum field theory.  Along the way we
will see that the contributions have been both practical and
fundamental.  They are practical in the sense that we can perform
quantitative computer calculations of non-perturbative effects in the
strong interactions.  They are fundamental in the sense that the
lattice gives deep insights into the workings of relativistic field
theory, in particular into anomalous features that distinguish between
the classical and the quantum theories.

\section {Before the lattice}

I begin by summarizing the situation in particle physics in the late
60's, when I was a graduate student.  Quantum-electrodynamics had
already been immensely successful, but that theory was in some sense
``done.''  While hard calculations remained, and indeed still remain,
there was no major conceptual advance remaining.

These were the years when the ``eightfold way'' for describing
multiplets of particles had recently gained widespread acceptance.
The idea of ``quarks'' was around, but with considerable caution about
assigning them any physical reality; maybe they were nothing but a
useful mathematical construct.  A few insightful theorists were
working on the weak interactions, and the basic electroweak
unification was beginning to emerge.  The SLAC experiments were
observing substantial inelastic electron-proton scattering at large
angles, and this was quickly interpreted as evidence for substructure,
with the term ``parton'' coming into play.  While occasionally there
were speculations relating quarks and partons, people tended to be
rather cautious about pushing this too hard.

A crucial feature of the time was that the extension of quantum
electrodynamics to a meson-nucleon field theory was failing miserably.
The analog of the electromagnetic coupling had a value about 15, in
comparison with the 1/137 of QED.  This meant that higher order
corrections to perturbative processes were substantially larger than
the initial calculations.  There was no known small parameter in which
to expand.

In frustration over this situation, much of the particle theory
community set aside traditional quantum field theoretical methods and
explored the possibility that particle interactions might be
completely determined by fundamental postulates such as analyticity
and unitarity.  This ``S-matrix'' approach raised the deep question of
just ``what is elementary?''  A delta baryon might be regarded as a
combination of a proton and a pion, but it would be just as correct to
regard the proton as a bound state of a pion with a delta.  All
particles are bound together by exchanging themselves.  These ``dual''
views of the basic objects of the theory persist today in string
theory.

\section{ The birth of QCD}

As we entered the 1970's, partons were increasingly identified with
quarks.  This shift was pushed by two dramatic theoretical
accomplishments.  First was the proof of renormalizability for
non-Abelian gauge theories\cite{renormalizability}, giving confidence
that these elegant mathematical structures\cite{ym} might have
something to do with reality.  Second was the discovery of asymptotic
freedom, the fact that interactions in Yang-Mills theories become
weaker at short distances\cite{asymptoticfreedom}.  Indeed, this was
quickly connected with the point-like structures hinted at in the SLAC
experiments.  Out of these ideas evolved QCD, the theory of quark
confining dynamics.

The viability of this picture depended upon the concept of
``confinement.''  While there was strong evidence for quark
substructure, no free quarks were ever observed.  This was
particularly puzzling given the nearly free nature of their apparent
interactions inside the nucleon.  This returns us to the question of
``what is elementary?''  Are the fundamental objects the physical
particles we see in the laboratory or are they these postulated quarks
and gluons?

\begin{figure}
\centering
\includegraphics[width=.7\hsize]{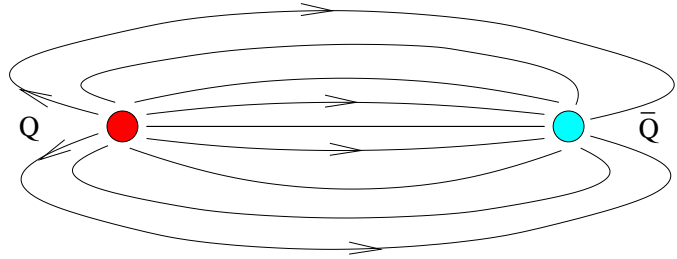}
\caption {A tube of gluonic flux connects quarks and anti-quarks.  The
strength of this string is 14 tons.}
\label{fluxtube}
\end{figure}

Struggling with this paradox led to the now standard flux-tube picture
of confinement.  The eight gluons are analogues of photons except that
they carry ``charge'' with respect to each other.  Without confinement
gluons would presumably be free massless particles like the photon.
But a massless charged particle would be a rather peculiar object.
Indeed, what happens to the self energy in the electric fields around
a gluon?  Such questions naturally lead to a conjectured instability
of the {\ae}ther that removes zero mass gluons from the spectrum.  This
is to be done in a way that does not violate Gauss's law.  Note that a
Coulombic $1/r^2$ field is a solution of the equations of a massless
field, not a massive one.  Without massless particles in the spectrum,
such a spreading of the gluonic flux is not allowed since it cannot
satisfy the appropriate equations in the weak field limit.  But from
Gauss's law, the field lines emanating from a quark cannot end.
Instead of spreading in the inverse square manner, the flux lines
cluster together, forming a tube emanating from the quark and
ultimately ending on an anti-quark as sketched in
Fig.~{\ref{fluxtube}}.  This structure is a real physical object, and
grows in length as the quark and anti-quark are pulled apart.  The
resulting force is constant at long distance, and is measured via the
spectrum of high angular momentum states, organized into the famous
``Regge trajectories.''  In physical units, the flux tube pulls with a
strength of about 14 tons.

The reason a quark cannot be isolated is similar to the reason that a
piece of string cannot have just one end.  Of course one can't have a
piece of string with three ends either, but this is the reason for the
underlying $SU(3)$ group theory, wherein three fundamental charges can
form a neutral singlet.  It is important to emphasize that the
confinement phenomenon cannot be seen in perturbation theory; when the
coupling is turned off, the spectrum becomes free quarks and gluons,
dramatically different than the pions and protons of the interacting
theory.

\section{ The 70's revolution}

The discoveries related to the Yang-Mills theory were just the
beginning of a revolutionary sequence of events in particle physics.
Perhaps the most dramatic was the discovery of the $J/\psi$
particle\cite{jpsi}.  The interpretation of this object and its
partners as bound states of heavy quarks provided the hydrogen atom of
QCD.  The idea of quarks became inescapable; field theory was reborn.
The $SU(3)$ non-Abelian gauge theory of the strong interactions was
combined with the recently developed electroweak theory to become the
durable ``standard model.''

This same period also witnessed several additional remarkable events
on the more theoretical front.  Non-linear effects in classical field
theories were shown to have deep consequences for their quantum
counterparts.  Classical ``lumps'' represented a new way to get
particles out of a quantum field theory\cite{lumps}.  Much of the
progress here was in two dimensions, where techniques such as
``bosonization'' showed equivalences between theories of drastically
different appearance.  A boson in one approach might appear as a bound
state of fermions in another, but in terms of the respective
Lagrangian approaches, they were equally fundamental.  Again, we were
faced with the question ``what is elementary?''  Of course modern
string theory is discovering multitudes of ``dualities'' that continue
to raise this same question.

The ensuing obsession with classical solutions quickly led to the
discovery \cite{Belavin:fg}
of ``pseudo-particles'' or ``instantons'' as classical
solutions of the four dimensional Yang-Mills theory in Euclidean space
time.  See R. Jackiw's contribution to this volume.  These turned out
to be intimately related to the famous anomalies in current algebra,
and gave a simple mechanism to generate the anomalous masses of such
particles as the $\eta^\prime$.  These effects were all inherently
non-perturbative, having an explicit exponential dependence in the
inverse coupling.  If the coupling is reduced in the theory with a
fixed cutoff, these effects fall to zero faster than any power of the
coupling.

This slew of discoveries had deep implications: field theory can
display much more structure than seen from the traditional analysis of
Feynman diagrams.  But this in turn had crucial consequences for
practical calculations.  Field theory is notorious for divergences
requiring regularization.  The bare mass and charge are divergent
quantities.  They are not the physical observables, which must be
defined in terms of physical processes.  To calculate, a ``regulator''
is required to tame the divergences, and when physical quantities are
related to each other, any regulator dependence should drop out.

The need for controlling infinities had, of course, been known since
the early days of QED.  But all regulators in common use were based on
Feynman diagrams; the theorist would calculate diagrams until one
diverged, and that diagram was then cut off.  Numerous schemes were
devised for this purpose, ranging from the Pauli-Villars approach to
forest formulae to dimensional regularization.  But with the
increasing realization that non-perturbative phenomena were crucial,
it was becoming clear that we needed a ``non-perturbative'' regulator,
independent of diagrams.

\section{The lattice}

The necessary tool appeared with Wilson's lattice theory.  He
originally presented this as an example of a model exhibiting
confinement.  The strong coupling expansion has a non-zero radius of
convergence, allowing a rigorous demonstration of confinement, albeit
in an unphysical limit.  The resulting spectrum has exactly the
desired properties; only gauge singlet bound states of quarks and
gluons can propagate.

This was not the first time that the basic structure of lattice gauge
theory had been written down.  A few years earlier,
Wegner\cite{wegner} presented a $Z_2$ lattice gauge model as an
example of a system possessing a phase transition but not exhibiting
any local order parameter.  In his thesis, Jan Smit\cite{smit}
described using a lattice regulator to formulate gauge theories
outside of perturbation theory.  The time was clearly ripe for the
development of such a regulator.  Very quickly after Wilson's
suggestion, Balian, Drouffe, and Itzykson\cite{bdi} explored an
amazingly wide variety of aspects of these models.

To reiterate, the primary role of the lattice is to provide a
non-perturbative cutoff.  Space is not really meant to be a crystal,
the lattice is a mathematical trick.  It provides a minimum wavelength
through the lattice spacing $a$, {\it i.e.} a maximum momentum of
$\pi/a$.  Path summations become well defined ordinary integrals.  By
avoiding the convergence difficulties of perturbation theory, the
lattice provides a route towards the rigorous definition of quantum
field theory.

The approach, however, had a marvelous side effect.  By discreetly
making the system discrete, it becomes sufficiently well defined to be
placed on a computer.  This was fairly straightforward, and came at
the same time that computers were growing rapidly in power.  Indeed,
numerical simulations and computer capabilities have continued to grow
together, making these efforts the mainstay of modern lattice gauge
theory.

\section{Gauge fields and phases}

As formulated by Wilson, the lattice cutoff is quite remarkable in
that it manages to keep exact many of the concepts of a gauge theory.
Of course, there are many ways to think of a gauge theory, and this is
apparent in the variety of viewpoints expressed in the contributions
to this volume.

At the most simplistic level, a Yang-Mills theory is just
electrodynamics embellished with isospin symmetry.  By working
directly with elements of the gauge group, this is inherent in lattice
gauge theory from the start.

At another level, a gauge theory is a theory of phases acquired by a
particle as it passes through space time.  Using group elements on
links directly gives this connection, with the phase associated with
some world-line being the product of these elements along the path in
question.  Of course, for the Yang-Mills theory the concept of
``phase'' becomes a rotation in the internal symmetry group.

A gauge theory is a theory with a local symmetry.  With the Wilson
action being formulated in terms of products of group elements around
closed loops, this symmetry remains exact even with the cutoff in place.  

In perturbative discussions, the local symmetry forces a gauge fixing
to remove a formal infinity of different gauges.  For the lattice
formulation, however, the use of a compact representation for the
group elements means that the integration over all gauges is finite.
To study gauge invariant observables, no gauge fixing is required to
define the theory.  Of course gauge fixing can still be done, and must
be introduced to study more conventional gauge variant quantities such
as gluon or quark propagators.

The only definition of a gauge theory that the lattice does not keep
exact is how a gauge field transforms under Lorentz transformations.
In a continuum theory the basic vector potential can change under a
gauge transformation when transforming between frames.  The lattice,
of course, breaks Lorentz invariance, and thus this concept looses
meaning.   
 
\section{ The Wilson action}

\begin{figure}
\centerline{
\includegraphics[width=.45\hsize]{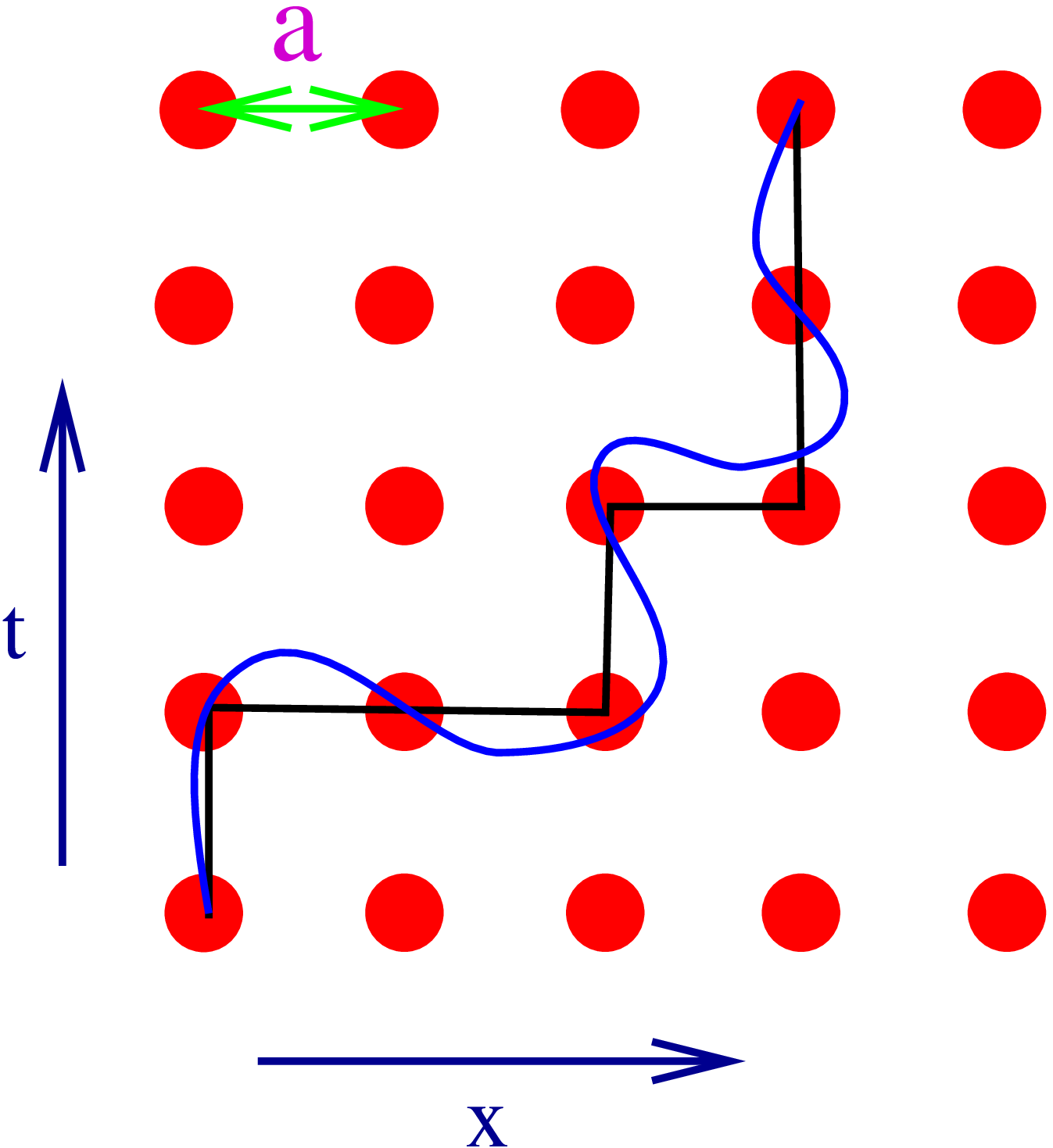}
}
\caption{In lattice gauge theory the world-line describing the motion
of a quark through space-time is approximated by a sequence of
discrete hops.  On each of these hops the quark wave function picks up
a ``phase'' described by the gauge fields.  For the strong
interactions, this phase is a unitary matrix in the group $SU(3)$.
}
\label{worldline}
\end{figure}

The concept of gauge fields as path dependent phases leads directly to
the conventional method for formulating the quark and gluon fields on
a lattice.  We approximate a general quark world-line by a set of
hoppings lying along lattice bonds, as sketched in Fig. \ref
{worldline}.  We then introduce the gauge field as group valued
matrices on these bonds.  Thus the gauge fields form a set of $SU(3)$
matrices, one such associated with every nearest neighbor bond on our
four-dimensional hyper-cubic lattice.

\begin{figure}
\centerline{
\includegraphics[width=.35\hsize]{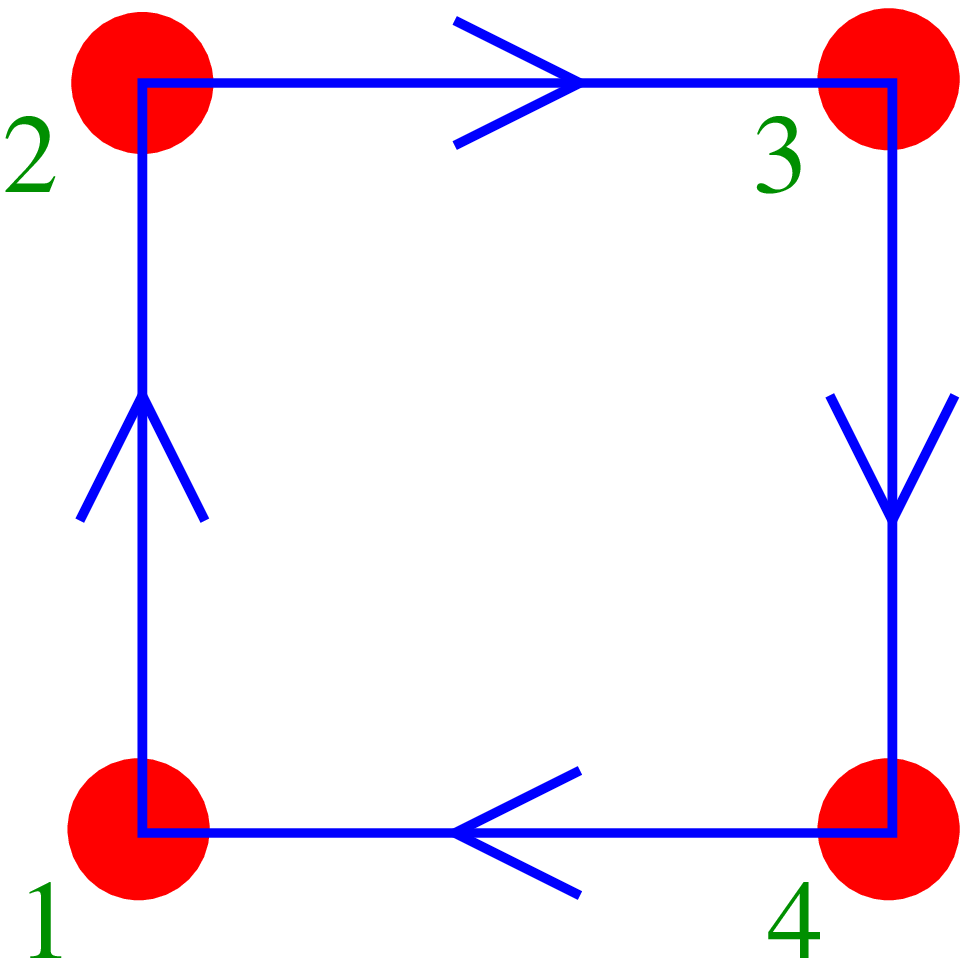}
}
\caption{In analogy with Stoke's law, the flux through an elementary
square of the lattice is found from the product of gauge matrices
around that square.  The dynamics is determined by adding the real
part of the trace of this product over all elementary squares.  This
``action'' is inserted into a ``path integral.''  The resulting
construction is formally a partition function for a system of
``spins'' existing in the group $SU(3)$.  }
\label{plaquette}
\end{figure}

In terms of these matrices, the gauge field dynamics takes a simple
natural form.  In analogy with regarding electromagnetic flux as the
generalized curl of the vector potential, we are led to identify the
flux through an elementary square, or ``plaquette,'' on the lattice
with the phase factor obtained on running around that plaquette; see
Fig. \ref {plaquette}.  Spatial plaquettes represent the ``magnetic''
effects and plaquettes with one direction time-like give the
``electric'' fields.  This motivates the conventional ``action'' used
for the gauge fields as a sum over all the elementary squares of the
lattice.  Around each square we multiply the phases and to get a real
number we take the real part of the trace
\begin{equation}
S_g=\sum_p {\rm Re\ Tr} \prod_{l\in p} U_l
\end{equation}  
Here the fundamental squares are denoted $p$ and the links $l$.  As we
are dealing with non-commuting matrices, the product around the square
is meant to be ordered.

To formulate the quantum theory of this system one usually uses the
Feynman path integral.  For this, exponentiate the action and
integrate over all dynamical variables to construct
\begin{equation}
Z=\int (dU) e^{-\beta S}
\end{equation}
where the parameter $\beta$ controls the bare coupling.  This converts
the three space dimensional quantum field theory of gluons into a
classical statistical mechanical system in four space-time dimensions.
Such a many-degree-of-freedom statistical system cries out for Monte
Carlo simulation, which now dominates the field of lattice QCD.  Note
the close analogy with a magnetic system; we could think of our
matrices as ``spins'' interacting through a four spin coupling
expressed in terms of the plaquettes.

The formulation is in Euclidean four dimensional space, based on an
underlying replacement of the time evolution operator $e^{-iHt}$ by
$e^{-Ht}$.  Despite involving the same Hamiltonian, excited states are
inherently suppressed and information on high energy scattering is
particularly hard to extract.  However low energy states and matrix
elements are the natural physical quantities to explore numerically.
This is the bread and butter of the lattice theorist.  Indeed, the
simulations reproduce the qualitative spectrum of stable hadrons quite
well.  Matrix elements currently under intense study are playing a
crucial role in ongoing tests of the standard model of particle
physics.

\section{A paucity of parameters}

Now I wish to reiterate one of the most remarkable aspects of the
theory of quarks and gluons, the small number of adjustable
parameters.  To begin with, the lattice spacing itself is not an
observable.  We are using the lattice to define the theory, and thus
for physics we are interested in the continuum limit $a\rightarrow 0$.
Then there is the coupling constant, which is also not a physical
parameter due to the phenomenon of asymptotic freedom.  The lattice
works directly with a bare coupling, and in the continuum limit this
should vanish as predicted by asymptotic freedom
\begin{equation}
g_0^2 \sim {1\over \log(1/\Lambda a)} \rightarrow 0
\end{equation} 
In the process, the coupling is replaced by an overall scale
$\Lambda$, which might be regarded as an integration constant for the
renormalization group equation.  Coleman and
Weinberg\cite{colemanweinberg} gave this phenomenon the marvelous name
``dimensional transmutation.''  Of course an overall scale is not
really something we should expect to calculate from first principles.
Its value would depend on the units chosen, be they furlongs or
light-fortnights.
 
Next consider the quark masses.  These also renormalize to zero as a
power of the coupling in the continuum limit.  Removing this
divergence, we can define a renormalized quark mass, which is a second
integration constant of the renormalization group equations.  One such
constant $M_i$ is needed for each quark ``flavor'' or species $i$.  Up
to an irrelevant overall scale, the physical theory is then a function
only of the dimensionless ratios $M_i/\Lambda$.  These are the only
free parameters in the strong interactions.  The origin of the
underlying masses remains one of the outstanding mysteries of particle
physics.

With multiple flavors, the massless quark limit gives a rather
remarkable theory, one with no undetermined dimensionless parameters.
This limit is not terribly far from reality; chiral symmetry breaking
should give massless pions, and experimentally the pion is
considerably lighter than the next non-strange hadron, the rho.  A
theory of two massless quarks is a fair approximation to the strong
interactions at intermediate energies.  In this limit all
dimensionless ratios should be calculable from first principles,
including quantities such as the rho to nucleon mass ratio.  The one
flavor theory provides an interesting intellectual exercise; indeed,
the massless one flavor theory is not uniquely
defined\cite{Creutz:2003xc}.

Since it is absorbed into an overall scale, the strong coupling
constant at any physical scale is not an input parameter, but should
be determined from first principles.  Such a calculation has gotten
lattice gauge theory into the famous particle data group
tables\cite{pdg}.  With appropriate definition a recent lattice result
is
\begin{equation}
\alpha_s(M_Z)=0.115\pm 0.003
\end{equation}
where the input is details of the charmonium spectrum.

\section{Numerical simulation}

While other techniques exist, such as strong coupling expansions,
large scale numerical simulations currently dominate lattice gauge
theory.  They are based on attempts to evaluate the path integral
\begin{equation}
Z=\int (dU)\ e^{-\beta S}
\end{equation}
with $\beta$ proportional to the inverse bare coupling squared.  A
direct evaluation of such an integral has pitfalls.  At first sight,
the basic size of the calculation is overwhelming.  Considering a
$10^4$ lattice, small by today standards, there are 40,000 links.  For
each is an $SU(3)$ matrix, parametrized by 8 numbers.  Thus we have a
$10^4\times 4 \times 8 = 320,000$ dimensional integral.  One might try
to replace this with a discrete sum over values of the integrand.  If
we make the extreme approximation of using only two points per
dimension, this gives a sum with
\begin{equation}
2^{320,000}=3.8\times 10^{96,329}
\end{equation}
terms!  Of course, computers are getting pretty fast, but one should
remember that the age of universe is only $\sim 10^{27}$ nanoseconds.

These huge numbers suggest a statistical treatment.  Indeed, the above
integral is formally just a partition function.  Consider a more
familiar statistical system, such as a glass of beer.  There are a
huge number of ways of arranging the atoms of carbon, hydrogen,
oxygen, etc.~that still leaves us with a glass of beer.  We don't need
to know all those arrangements, we only need a dozen or so ``typical''
glasses to know all the important properties.

This is the basis of the Monte Carlo approach.  The analogy with a
partition function and the role of ${1\over \beta}$ as a temperature
enables the use of standard techniques to obtain ``typical''
equilibrium configurations, where the probability of any given
configuration is given by the Boltzmann weight
\begin{equation}
P(C)\sim e^{-\beta S(C)}
\end{equation}
For this we use a Markov process, making changes in the current
configuration
\begin{equation}
C\rightarrow C^\prime \rightarrow \ldots
\end{equation}
biased by the desired weight.

The idea is easily demonstrated with the example of $Z_2$ lattice
gauge theory\cite{creutzjacobsrebbi}.  For this toy model the links
are allowed to take only two values, either plus or minus unity.
One sets up a loop over the lattice variables.  When looking at a
particular link, calculate the probability for it to have value $1$
\begin{equation}
P(1)={e^{-\beta S(1)}\over e^{-\beta S(1)}+e^{-\beta S(-1)}}
\end{equation} 
Then pull out a roulette wheel and select either 1 or $-1$ biased by
this weight.  Lattice gauge Monte-Carlo programs are by nature quite
simple.  They are basically a set of nested loops surrounding a random
change of the fundamental variables.

Extending this to fields in larger manifolds, such as the $SU(3)$
matrices representing the gluon fields, is straightforward.  The
algorithms are usually based on a detailed balance condition for a
local change of fields taking configuration $C$ to configuration
$C^\prime$.  If probabilities for making these changes in one step satisfy
\begin{equation}
{P(C\rightarrow C^\prime)\over P(C^\prime\rightarrow C)}
={e^{-\beta S(C^\prime)}\over e^{-\beta S(C)}}
\end{equation}
it is straightforward to prove that any ensemble of configurations
approaches the equilibrium ensemble.

The results of these simulations have been fantastic, giving first
principles calculations of interacting quantum field theory.  I will
just mention two examples.  The early result that bolstered the
lattice into mainstream particle physics was the convincing
demonstration of the confinement phenomenon.  The force between two
quark sources indeed remains constant at large distances.

Another accomplishment for which the lattice excels over all other
methods has been the study the deconfinement of quarks and gluons into
a plasma at a temperature of about 170--190 Mev\cite{plasma}.  Indeed,
the lattice is a unique quantitative tool capable of making precise
predictions for this temperature.  The method is based on the fact
that the Euclidean path integral in a finite temporal box directly
gives the physical finite temperature partition function, where the
size of the box is proportional to the inverse temperature.  This
transition represents the confining flux tubes becoming lost in a
background plasma of virtual flux lines.

\section{Quarks and random numbers}

While the gauge sector of the lattice theory is in good shape, from
the earliest days fermionic fields have caused annoying difficulties.
Actually there are several apparently unrelated fermion problems.  The
first is an algorithmic one.  The quark operators are not ordinary
numbers, but anti-commuting operators in a Grassmann space.  As such,
the exponentiated action itself is an operator.  This makes comparison
with random numbers problematic.

Until relatively recently, most lattice work with quarks was done in
the so called ``valence'' or ``quenched'' approximation.  A pure gauge
simulation provides a set of background gauge fields in which the
propagation of quarks is calculated.  The approximation is to ignore
any feedback of the quarks on the gauge fields.  As the quarks involve
large sparse matrices, the conjugate gradient algorithm is ideally
suited.  Combining the resulting propagators into hadronic
combinations gives predictions on physical quantities such as spectra,
matrix elements, etc.  The rather random nature of the relevant
background fields has hampered application of standard multi-scale
techniques, but more work in this area is needed.  The main issue with
the valence approximation is that systematic errors are not under
precise control.

Over the years various clever tricks for dealing with dynamical quarks
have been developed; numerous ongoing large scale Monte Carlo
simulations do involve dynamical fermions.  The algorithms used are
all essentially based on an initial analytic integration of the quarks
to give a determinant.  This, however, is the determinant of a rather
large matrix, the size being the number of lattice sites times the
number of fermion field components, with the latter including spinor,
flavor, and color factors.  For a Monte Carlo evolution we need to
know how this determinant changes with random changes in the gauge
field.  Introducing auxiliary bosonic fields reduces the problem to
doing large sparse matrix inversions inside the Monte Carlo loop.  It
is these inversions that currently dominate the required compute time.
In my opinion, the algorithms working directly with these large
matrices remain quite awkward.  I often wonder if there is some more
direct way to treat fermions without the initial analytic integration.
On small systems direct evaluation of Grassmann integrals by machine
is possible\cite{Creutz:1998ee,Creutz:2002tu}, although the approach
appears to be inherently exponential in the system volume.

The algorithmic problem becomes considerably more serious when a
chemical potential generating a background baryon density is present.
In this case the required determinant is not positive; it cannot be
incorporated as a weight in a Monte Carlo procedure.  This is
particularly frustrating in the light of striking predictions of
super-conducting phases at large chemical
potential\cite{superconduct}.  This is perhaps the most serious
unsolved problem in lattice gauge theory today.

\section{Chirality, anomalies, and the lattice}

While the difficulty in simulating Grassmann dynamics is a major issue,
further conceptual fermion problems concern chiral issues.  These are
intimately entwined with the anomalous differences between
classical and quantum field theories.  Indeed, while the lattice is
usually just thought of as a numerical technique, it also provides a
path to understanding many subtleties of quantum field theory.  As a
full non-perturbative regulator, the lattice provides a foundation
for defining quantum field theory.  

It is well known that some classical symmetries do not survive
quantization.  The most basic example, the scale anomaly, has been so
fully absorbed into the lattice lore that it is rarely mentioned.
The classical Yang-Mills theory is scale invariant and depends in a
non-trivial way on the coupling constant.  The quantum theory,
however, is not at all scale invariant.  Indeed, it is a theory of
massive glueballs and the masses these particles set a definite scale.

When the quark masses vanish, the classical Lagrangian for the strong
interactions still contains no dimensional parameters.  But the
quantum theory is supposed to describe baryons and mesons, and the
lightest baryon, the proton, definitely has mass.  As discussed in the
earlier section on parameters, this is understood through the
phenomenon of ``dimensional transmutation,'' wherein the classical
coupling constant of the theory is traded, through the process of
renormalization, for an overall scale parameter\cite{colemanweinberg}.

The scale anomaly is perhaps the deepest, but it is not the only
symmetry of the strong interactions of massless quarks that is lost
upon quantization.  The most famous are the anomalies in the
axial-vector fermion currents\cite{Bell:ts,Adler:gk,Adler:er}, also
discussed in the contributions by S. Adler and R. Jackiw to this
volume.  Working in a helicity basis, the classical Lagrangian has no
terms to change the number of left or right handed fermions.  On
quantization, however, these numbers cease to be separately conserved.
Technically this comes about because of the famous triangle diagram.
This introduces a divergence which requires regularization via a
dimensionful cutoff.  For the strong interactions with its vector-like
gluon couplings, this regulation is implemented so that the vector
current, representing total fermion number, is conserved.  But if this
choice is taken, then the axial current, representing the difference
of right and left handed fermion numbers, cannot be.  There is a
freedom in choosing which currents are conserved; however, in a gauge
theory, consistency requires that gauge fields couple only to
conserved currents.

In the full standard model, anomalies require some time honored
conservation laws to be violated.  The most famous example is baryon
number, which in the standard model is sacrificed so that the chiral
currents that couple to the vector bosons are
conserved\cite{'tHooft:up,'tHooft:fv}.  Baryon violating
semi-classical processes have been identified and must be present,
although at a very low rate.  While not of observable strength, at a
conceptual level any scheme for non-perturbatively regulating the
standard model must either contain baryon violating
terms\cite{Eichten:1985ft} or extend the model to cancel these
anomalies with, say, mirror
species\cite{Montvay:1987ys,Montvay:1992eg}.

Consistency under anomalies has non-trivial implications for the
allowed species of fermions.  To conserve all the gauged currents of
the standard model requires the cancellation of all potential
anomalies in currents coupled to gauge fields.  In particular, the
standard model is not consistent if either the leptons or the quarks
are left out.  This connection between quarks and leptons is a deep
subtlety of the theory and must play a key role in placing the theory
on a lattice.  Although these effects are extremely tiny due to the
smallness of the weak coupling constant, without a precise
non-perturbative regulator that is capapable of including these
phenomena, it is not clear that the weak interactions fit into a
meaningful field theory.

At a more phenomenological level, there are a variety of reasons that
chiral symmetries are important to particle physics.  Premier among
these is the light nature of the pion, which is traditionally related
to the spontaneous breaking of a chiral symmetry expected to become
exact as the quark masses go to zero.  This is the explanation as to
why the pion is so much lighter than the rho meson, even though they
are made of the same quarks, albeit in different spin states.

Theories unifying the various interactions also often make heavy use
of chiral symmetry.  Indeed, chiral symmetry protects fermion masses
from large renormalizations, helping control an unwanted generation of
large masses requiring fine tuning to avoid.  This is also one of the
main arguments for super-symmetry, enabling protection mechanisms for
bosonic masses such as that of the Higgs boson.

Despite its clear importance, chiral symmetry and the lattice have
never fit particularly well together.  When the lattice is in place,
there are no divergences.  Thus any symmetries of the defining action
must remain exact.  If we ignore the known anomalies in formulating
our actions, something must go wrong.  Indeed, the most naive methods
for including fermions have what is known as the ``doubling'' problem.
Extra species appear involving momentum components near the cutoff,
and including them makes the naive axial symmetry actually a vector
symmetry.  The doubling problem is not a nemesis, but a sign that the
lattice is trying to tell us something deep.

\begin{figure}
\centering
\includegraphics[width=.7\hsize]{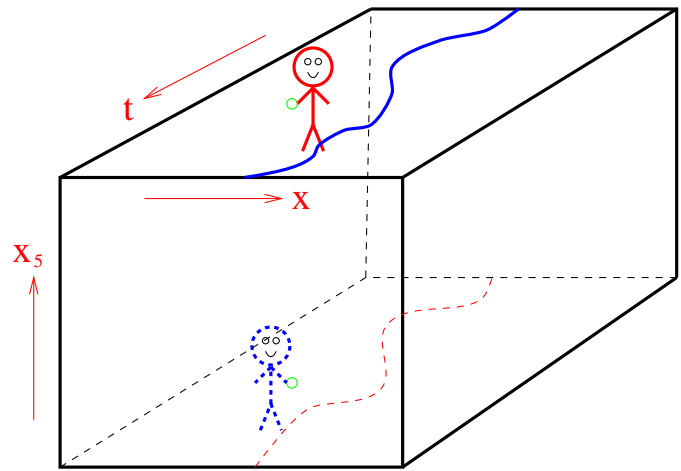}
\caption {In the domain wall approach we start with a five dimensional
lattice theory set up so that low energy fermionic states are bound to
the four dimensional surfaces.  Our four dimensional world arises as
energy required to create excitations traveling into the fifth
dimension goes to infinity.}
\label{kaplan}
\end{figure}

These issues are currently a topic with lots of activity.  For a
recent review, see\cite{myreview}.  This is not an appropriate place
to get involved in technical details, some of which remain unresolved.
Several elegant schemes for making chiral symmetry more manifest have
recently been developed.  My current favorite is the ``domain-wall''
formulation\cite{Kaplan:1992bt}, where our four dimensional world is
an interface in an underlying five dimensional theory, as sketched in
Fig.~(\ref{kaplan}).  The five dimensional quarks are given masses of
order the cutoff, but the basic action is adjusted so that there are
topologically stable zero mass modes bound on the surfaces of the
system.  At low energies in the continuum limit only these four
dimensional modes are excited.
  
This approach works quite well for vector-like theories, with opposite
chirality quarks living on opposite walls of the five dimensional
theory.  For chiral gauge theories, however, it is necessary to
eliminate the modes on one of the two walls.  It is not known how to
do this in a clean way since the gauge fields do not know about the
fifth dimension, and thus see both walls.  Various techniques have
been proposed to give a large mass to excitations on the the unwanted
wall.  This could be done with a Higgs coupling that becomes large on
one wall; this is effectively a mirror fermion model.  Another
proposal involves artificially increasing the strength of the 't Hooft
vertex on the unwanted wall\cite{Creutz:1996xc}; this involves four
fermion couplings at the scale of the cutoff and is very difficult to
treat rigorously.

Closely related to the domain wall approach are the ``Ginsparg-Wilson'
fermionic actions, which maintain an exact, albeit somewhat more
complicated, chiral symmetry\cite{overlap,admissible,gw}.  This
approach is mathematically extremely elegant, giving rise to an exact
lattice version of the continuum index theorem relating zero
eigenvalues of the Dirac operator with the topological index of the
gauge fields.  While a lattice regularization of a full chiral gauge
theory such as the standard model remains elusive, we may not be far
off.

\section{Concluding remarks}

In summary, lattice gauge theory provides the dominant framework for
investigating non-perturbative phenomena in quantum field theory.  The
approach is currently dominated by numerical simulations, although the
basic framework is potentially considerably more flexible.  With the
recent developments towards implementing chiral symmetry on the
lattice, including domain-wall fermions, the overlap formula, and
variants on the Ginsparg-Wilson relation, parity conserving theories,
such as the strong interactions, are fundamentally in quite good
shape.

I personally am fascinated by the chiral gauge problem.  Without a
proper lattice formulation of a chiral gauge theory, it is unclear
whether such models make any sense as a fundamental field theories.  A
marvelous goal would be a fully finite, gauge invariant, and local
lattice formulation of the standard model.  The problems encountered
with chiral gauge theory are closely related to similar issues with
super-symmetry, another area that does not naturally fit on the
lattice.  This also ties in with the explosive activity in string
theory and a possible regularization of gravity.

The other major unsolved problems in lattice gauge theory are
algorithmic.  Current fermion algorithms are extremely awkward and
computer intensive.  It is unclear why this has to be so, and may only
be a consequence of our working directly with fermion determinants.
One could to this for bosons too, but that would clearly be terribly
inefficient.  At present, the fermion problem seems completely
intractable when the fermion determinant is not positive.  This is of
more than academic interest since interesting superconducting phases
are predicted at high quark density.  Similar sign problems appear in
other fields, such as doped strongly coupled electron systems, thus
making this problem practically quite important.

Finally, throughout history the question of ``what is elementary?''
continues to arise.  This is almost certainly an ill posed question,
with one or another approach being simpler in the appropriate context.
See E. Witten's contribution to this volume for a discussion of some
of the modern equivalences.  At a more mundane level, for low energy
chiral dynamics we lose nothing by considering the pion as an
elementary pseudo-goldstone field, while at extremely short distances
string structures may become more fundamental.  Quarks and their
confinement may just be useful temporary constructs along the way.

\section*{Acknowledgment}{This manuscript has been authored under
contract number DE-AC02-98CH10886 with the U.S.~Department of Energy.
Accordingly, the U.S. Government retains a non-exclusive, royalty-free
license to publish or reproduce the published form of this
contribution, or allow others to do so, for U.S.~Government purposes.}

\end{document}